**Brief increases in corticosterone affect morphology, stress responses, and telomere length, but not post-fledging movements, in a wild songbird**

**Running head: Brief developmental corticosterone doses affect phenotype**

Teresa M. Pegan[1,3]*, David W. Winkler[1], Mark F. Haussmann[2], and Maren N. Vitousek[1]

[1]Department of Ecology and Evolutionary Biology, Cornell University, Ithaca, New York, USA

[2]Department of Biology, Bucknell University, Lewisburg, Pennsylvania, USA

[3]Current address: Department of Ecology and Evolutionary Biology, University of Michigan, Ann Arbor, Michigan, USA

*Corresponding author: tmpegan@umich.edu

What is already known: Experimental manipulations of glucocorticoids during development affect various phenotypic traits, including morphology, HPA axis functioning, and behavior. Most such studies have manipulated glucocorticoids to mimic an increase in baseline glucocorticoid levels, rather than an acute stress response. Because most manipulations have been done in captive populations, impacts of glucocorticoids on phenotype, movement patterns, and fitness under natural conditions are not well understood.

What this study adds: Using a wild, free-living songbird, we demonstrate that just five brief, acute increases in the glucocorticoid stress hormone corticosterone during development are sufficient to produce phenotypic effects including reduced mass, higher baseline and stress-induced corticosterone levels, and shorter telomeres. We did not find evidence that this manipulation affected time to maturity, post-fledging behavior, or survival.

**Abstract**

Organisms are frequently exposed to challenges during development, such as poor weather and food shortage. Such challenges can initiate the hormonal stress response, which involves secretion of glucocorticoids. Although the hormonal stress response helps organisms deal with challenges, long-term exposure to high levels of glucocorticoids can have morphological, behavioral, and physiological consequences, especially during development. Glucocorticoids are also associated with reduced survival and telomere shortening. To investigate whether brief, acute exposures to glucocorticoids can also produce these phenotypic effects in free-living birds, we exposed wild tree swallow (*Tachycineta bicolor*) nestlings to a brief exogenous dose of cort once per day for five days and then measured their morphology, baseline and stress-induced corticosterone levels, and telomere length. We also deployed radio tags on a subset of nestlings, which allowed us to determine the age at which tagged nestlings left the nest (fledged) and their pattern of presence and absence at the natal site during the post-breeding period. Corticosterone-treated nestlings had lower mass, higher baseline and stress-induced corticosterone, and reduced telomeres; other metrics of morphology were affected weakly or not at all. Our treatment resulted in no significant effect on survival to fledging, fledge age, or age at first departure from the natal site, and we found no negative effect of corticosterone on inter-annual return rate. These results show that brief acute corticosterone exposure during development can have measurable effects on phenotype in free-living tree swallows. Corticosterone may therefore mediate correlations between rearing environment and phenotype in developing organisms, even in the absence of prolonged stressors.

**Introduction**

Organisms are frequently exposed to a variety of stressors, and their responses to these challenges are generally mediated by hormones through the glucocorticoid stress response (McEwan and Wingfield 2003). The low, variable level of glucocorticoid hormones circulating in an organism's bloodstream in the absence of a stressor helps allocate energy in the face of predictable demands; after exposure to a stressor, organisms experience a rapid increase in circulating glucocorticoids (the stress response), which mediates behavioral and physiological responses that help to maintain homeostasis (McEwan and Wingfield 2003, Landys et al. 2006). At the same time, acute increases in glucocorticoids can inhibit immune function, reproductive behavior, and growth (Sapolsky et al. 2000).

Exposure to stress hormones can have effects on phenotype that persist after the stressor has ceased. This may be especially true during development, when hormones can have organizational as well as activational effects (Schoech et al. 2011). Many vertebrates, including birds, spend much of their developmental period exposed to the environment and at least partially dependent on their parents for care. During this time, the glucocorticoid stress response can be activated in response to both acute and chronic challenges (Schoech et al. 2011). Studies in captive birds have shown that acute glucocorticoid exposure during development can suppress mass gain (Loiseau et al. 2008a; Spencer and Verhulst 2007; Crino et al. 2014; but see Loiseau et al. 2008b), alter immune function (Loiseau et al. 2008a, 2008b) and metabolic rate (Spencer and Verhulst 2008), change nestling begging behavior (Loiseau et al. 2008b, Wada and Breuner 2008), and influence the glucocorticoid response to future stressors (Haussman et al. 2012, Schmidt et al. 2014). At least some of these phenotypic effects can persist into adulthood (Crino et al. 2014, but see Spencer and Verhulst 2008). The majority of studies on the relationship between phenotype and stress exposure during development have occurred in lab-reared organisms.

Exposure to stressors or to high glucocorticoid levels during development may also impact telomere length (Haussmann et al. 2012, Boonekamp et al. 2014, Herborn et al. 2014). Telomeres are noncoding regions at the ends of eukaryote chromosomes that protect DNA from damage. Telomeres progressively shorten with each cell division throughout an animal's life, and they can also shorten as a result of damage by oxidative stress, which in turn can be exacerbated by glucocorticoids (reviewed in Haussmann and Marchetto 2010, but see Boonekamp et al. 2017). Studies have linked telomere shortening with reduced survival (Haussmann et al. 2005, Bakaysa et al 2007, Boonekamp et al. 2014). In this way, developmental stress could have a far-reaching impact on the subsequent life of the animal. However, the relationship between glucocorticoids and survival is likely complex and context-dependent: elevated glucocorticoids have also been associated with increased survival (Rivers et al 2012).

One of the behavioral consequences of increased glucocorticoid concentration is increased locomotor activity (Challet et al. 1995, Breuner et al. 1998, Vercken et al. 2007, Meylan et al. 2009), suggesting that these hormones may have an important effect on movement behaviors. Glucocorticoids are known to spike immediately prior to important transitions in organisms' lives involving movement, including departure from the nest (called "fledging" in birds; Sprague and Breuner 2010) and departure from the natal territory to disperse or explore the surrounding area (Belthoff and Dufty 1998). Experimentally elevating the glucocorticoid hormone corticosterone in already-independent juvenile birds can prompt them to disperse away from their natal territories (Silverin 1997). If the glucocorticoid level experienced during early development affects later life transitions and movement behaviors, these hormones could serve as a physiological link between the conditions experienced during development and adaptive movement decisions. However, the effect of stress experienced earlier in development on later movement behavior is unclear. For example, although developmental stress can cause long-term increases in HPA activity (Schmidt et al. 2014), potentially leading to earlier fledging as a result of high glucocorticoid levels, elevated glucocorticoids also slow physical development (Hayward and Wingfield 2004, Loiseau et al. 2008a; Spencer and Verhulst 2007; Crino et al. 2014), which may delay departure from the nest (Michaud and Leonard 2000).

In this study, we investigate the effects of brief, acute increases in the glucocorticoid hormone corticosterone on phenotype in developing tree swallows *(Tachycineta bicolor),* a species which has long served as a model organism for wild songbirds (Jones 2003). Manipulating corticosterone in a free-living organism allowed us to test whether brief experimental increases in this hormone would have a persistent effect on phenotype in a natural setting. By applying radio tags to a subset of the individuals in our experiment, we were also able to test whether elevated corticosterone early in development would affect behaviors that are impossible to observe in lab settings, including age at initial departure from the natal site and the amount of time spent at the natal site between fledging and embarking on fall migration (a metric which may be associated with prospecting for future nesting sites: Morton 1992, Reed et al 1999). We predicted that nestlings exposed to a series of brief, acute increases in corticosterone would grow more slowly and would show altered stress physiology (higher baseline and stress-induced cort), shorter telomeres, and reduced survival. We also predicted that they would fledge earlier and depart from the natal site earlier. Because we expected individuals with higher corticosterone levels to have higher motivation to disperse away from the natal site instead of returning there to breed, we predicted that they would show reduced presence at natal site between fledging and the onset of fall migration.

**Methods**

*Organism and Experimental Site*

Tree swallows are small (~20 g) migratory songbirds that breed across North America and spend the winter in southern North America and the Caribbean. We conducted our study at a colony of tree swallows nesting in boxes near a system of ponds in Ithaca, NY, during the summer of 2015. Although the nest boxes are provided by researchers, the swallows living in them are entirely free-living. This colony (which we will refer to as the "natal site") has been monitored for over 30 years; details about the site can be found in Winkler and Allen (1996). Tree swallow nestlings in this colony may face a number of stressors during development, including predation attempts and the challenges associated with periods of cold weather, which can cause widespread juvenile mortality by reducing the number of flying insects (this species' primary food source; Winkler et al 2013). Nestlings respond to stressors, including cold periods, with a rapid increase in circulating corticosterone levels (Vitousek et al., unpublished data).

The eggs in all nests used in this experiment were laid between May 24 and June 5, 2015. We monitored the nestlings until fledging (the time at which they leave the nest and become independent). A subset of nestlings were radio-tagged prior to fledging, allowing us to track their presence or absence at the natal site until the onset of their fall migration.

*Corticosterone Manipulation*

We applied a corticosterone treatment to half of the nestlings in each of 21 experimental nests, leaving the other half as controls. On what we estimated to be the fifth day after nestlings hatched ("day 5") we began the experimental treatment. After weighing the nestlings, we randomly assigned them to treatment groups so that within each nest, each experimental group contained approximately the same number of nestlings. We uniquely identified each nestling by coloring the flesh near its beak (the gape) with Prismacolor non-toxic permanent markers (Prismacolor, Oak Brook, Illinois, USA; see e.g., Safran et al. 2005).

We applied treatments topically to apteria (featherless portions of the skin) using a 1 ml syringe with no needle. Experimental birds received corticosterone dissolved in dimethyl sulfoxide (DMSO) gel at a concentration of $0.5\mu g\ \mu l^{-1}$. In adults, a similar method raises corticosterone for ~60-90 mins to elevated levels that are within the range of natural variation (Vitousek et al., 2018). Nestlings in the control group received the vehicle (DMSO) alone. The nestlings received one dose of either DMSO or corticosterone-infused DMSO per day from day 5 to day 9. We increased the doses as the nestlings grew to give each nestling roughly 2µl of gel (and thus 1µg corticosterone in the treatment group) per gram of bird. The dosing schedule was as follows: 20µl (day 5), 20µl (day 6), 25µl (day 7), 30µl (day 8), 30µl (day 9). To minimize handling stress, we applied the doses without removing nestlings from the nest. We varied the time at which we applied the doses so that nestlings would not become acclimated to the doses.

On nestling day 12, we measured the length of the skull from the back of the head to the bill tip (head-bill), flat wing length, and body mass. Between 1100 hours and 1400 hours, we took two blood samples from each nestling via brachial venipuncture. The first occurred within 3 minutes of disturbance (allowing us to measure baseline, pre-disturbance corticosterone level) and the second occurred 30 minutes later (allowing us to measure the stress-induced change in circulating corticosterone level). The mean amount of blood taken from each nestling (for both samples combined) was ~100µl and the maximum amount was ~190µl. We stored blood samples on ice and later spun them at 3500 rpm for 5 min to separate plasma from red blood cells. We stored plasma at -30 ºC, and we stored red blood cells in lysis buffer. For some individuals, we also stored a portion of red blood cells in calf serum buffer at -70 ºC for subsequent telomere analysis.

*Radio Tagging*

Between day 15 and day 17, we radio-tagged the heaviest nestling from each experimental group in each nest (two nestlings per nest except at one nest, which had only one nestling tagged); 20 treatment birds and 21 control birds were tagged. We used solar-powered Life Tags (Cellular Tracking Technologies, Cape May NJ) and attached the tags with a leg loop harness (Rappole and Tipton 1991). Ultimately, 36 nestlings fledged with a tag. Five nestlings with tags (~12%) died before fledging; 22% of non-tagged birds from experimental nests that lived to day 12 also died before fledging. We selected heavy nestlings for tagging because heavy nestlings are more likely to be recaptured in subsequent years (McCarty 2001), but doing so also inadvertently biased our sample of tagged birds toward male nestlings. Of the 36 nestlings that fledged with a radio tag, 29 (80.6%) were males.

Life Tags use solar cells rather than batteries, and thus do not have a prescribed limit to the amount of time they are functional. These early-generation Life Tags lasted less than one year before ceasing to function due to a problem with the attachment between the tag and the antenna transmitting its radio signals (this problem has been fixed in the current generation of Life Tags; Winkler, unpublished data). We suspect that this failure occurred at a wide scale during the early months of 2016 - and that our data from the fall of 2015 are minimally affected by tag failure - because we successfully received signals from tagged adult birds (not involved in this study) in the early spring of 2016, but ceased picking up signals and began encountering birds with broken tags in May and June of 2016 (Pegan et al, unpublished data).

We used an autonomous receiver with a 1.67-m Diamond X50NA omnidirectional antenna (Diamond Antenna, San Marcos, CA) to collect data on the presence of tagged juveniles at the natal site between 7 June 2015 and 25 October 2015. We considered the "natal site" to be a circular area with a radius of 500 m centered over the tree swallow colony. This circle, which also represents the total area over which tags could be detected by our autonomous receiver, includes all of the nest boxes at the site (which were all < 400 m from the receiver) and a small amount of surrounding area; the receiver was placed at approximately the center of the circle. During the time the receiver was present at the site, it was nonfunctional for ~18 days (13% of total time) because of battery discharging and mechanical issues. For an additional 23 days (16% of total time), it was either nonfunctional for part of the day, or its operational status could not be confirmed. All but three of these off-days occurred in August, after all of the nestlings had already fledged and made their first departures from the natal site; but it is possible that some late-season natal site visits were missed while the base station was nonfunctional.

*Corticosterone Assays*

We assayed corticosterone levels using DetectX Enzyme Immunoassay kits (Arbor Assays, Ann Arbor, Michigan, USA: K014-HB) after first extracting steroids from plasma using a triple ethyl acetate extraction (average extraction efficiency 85.4%; Stedman et al 2017). We ran samples in duplicate along with a 9-standard curve. In the few cases for which samples had corticosterone concentrations below the detection threshold (0.8 ng mL$^{-1}$), we used the detection threshold as the corticosterone concentration for those samples. Intra-assay variation was 5.4%, and inter-assay variation was 5.7% .

*Telomere Analyses*

We quantified telomere length using an avian-specific southern hybridization telomere restriction fragment assay technique (Haussmann et al. 2003, Haussmann et al. 2012). After extraction, DNA was restriction digested and run by pulsed field gel electrophoresis (3 V cm$^{-1}$, 0.5–7.0 s switch times, 14 ºC) for 19 h. Following this, the samples were hybridized within the gel at 37 ºC overnight with a $^{32}$P-labeled oligo (CCCTAA)$_4$. We scanned hybridized gels on phosphorscreens using a Storm 540 Variable Mode Imager (Amersham BioSciences, Little Chalfont, United Kingdom). We used densitometry (IMAGEQUANT 5.03; and IMAGEJ 1.42q, Schneider et al. 2012) to determine the position and strength of the $^{32}$P-signal in each lane compared with the molecular marker (1 kb DNA Extension Ladder; Invitrogen, Carlsbad, California, USA). We collected information on the entire telomere restriction fragment (TRF) distribution, and frequency distributions of telomere length were created for each individual (Kimura et al. 2008, Haussmann et al. 2012).

*Genetic Sexing*

We sexed the nestlings using DNA extracted from red blood cells by using a P2/P8 protocol with a HaeIII digest (Whittingham and Dunn 2000). We performed PCR using 1 μl 10x PCR buffer, 0.60 μl MgCl2, 1.3 μl of each primer, 0.2 μl dNTPs, 0.10 μl Taq polymerase, 2 μl DNA, and enough nuclease-free water to bring the final reaction volume to 10 μl. PCR conditions were as follows: an initial denaturation step of 1 min at 94 °C, followed by 34 cycles of 30 s of denaturation at 94 °C, 45 s of annealing at 46 °C, and 45 s of extension at 72 °C, and finally, 2 cycles of 30 s at 94 °C, 45 s at 47 °C, and 5 min at 72 °C. The HaeIII restriction enzyme digest is necessary in this species because in this species, the Z- and W-specific alleles of the CHD1 gene amplified by the P2/P8 primers are too similar in size to visually distinguish in a gel. Because a HaeIII cutting site is present on the CHD1-Z gene, but not on the CHD1-W gene (Griffiths et al. 1998), this digestion shortens only the Z-specific allele, enabling visual differentiation. Each digest consisted of 1 µl nuclease-free water, 1 µl 10x PCR buffer, 1 µl of restriction enzyme, and 7 µl PCR product. Digests were incubated for 3 hours at 37 °C, followed by an inactivation step of 20 min at 80 °C, after which PCR products were visualized using gel electrophoresis.

*Data Analyses*

We analyzed our data using R 3.4.1 (R Core Team 2017). Using linear mixed effect models, we tested the effect of corticosterone treatment on the following developmental and physiological response variables, all measurements taken on the nestlings' day 12: mass, head-bill length, wing length, scaled body mass (a measure of condition calculated based on mass and head-bill length, following Peig and Green 2009), baseline corticosterone, stress-induced corticosterone, and telomere length distribution. Using the subset of radio-tagged individuals, we used linear mixed-effect models to test: the effect of corticosterone treatment on the age at which individuals fledged (departed from the nest); the age at which individuals departed the natal site for the first time; and the total number of days individuals spent at the natal site after fledging. The latter variable was only weakly correlated with the age at first departure from the natal site (Pearson's product-moment correlation estimate = 0.20, $P$ = 0.24) because many individuals returned to the natal site repeatedly after departing it for the first time. Finally, we used generalized linear mixed models with a binomial distribution to test the effect of corticosterone treatment on survival to fledging, and the probability that an individual would survive and return to the study site the following breeding season. Inter-annual return was noted when birds were captured during the following year in mist nets or nest boxes at the natal site or – in one case – at a site 2 km away, where a bird's radio-tag signal was detected. Note, however, that a failure to return does not necessarily indicate mortality; young tree swallows, particularly females, often disperse some distance from their natal site (Winkler et al. 2005).

For each response variable, we tested 6 candidate models, all of which included nest identity as a random factor. Four models included treatment, either alone or in combination with other variables often correlated with developmental and physiological traits: treatment, treatment + sex, treatment + sex + treatment:sex, treatment + sex + brood size + hatch date. To test whether models that excluded treatment performed better, 2 additional models were tested: sex + brood size + hatch date, and a null model (random effect alone). Because nestlings were sexed with blood taken on day 12 and some nestlings died before day 12, we could not use sex as a predictor for survival to fledging. For analyses of survival to fledging and inter-annual return rate, we accounted for the effect of radio tagging by including a binary predictor indicating whether the individual was radio tagged in each model except for the null model. This variable was irrelevant for all other analyses because all other response variables because these were either measured before radio tagging (all response variables measured on day 12), or could be tested only with radio-tagged individuals (age at fledging, age at first departure from the natal site, and total number of days spent at the site). We compared the candidate models using the Akaike Information Criterion (AICc). Full model selection results are presented in the appendix, available online.

All candidate models for a given response variable included the same subset of data, but models for different response variables sometimes differed in sample size. We excluded two nests from all analyses of morphology and HPA activity because they were older than 12 days old when measurements were taken. Nestlings that died before we took phenotype measurements on day 12 are only included in models predicting survival to fledging. We excluded one nest that was taken by a predator late in the season from analysis of survival from day 5 to fledging. Sample sizes for physiological tests vary, as were not able to obtain sufficient plasma from all individuals to assess stress-induced samples ($n$ = 110 individuals with baseline corticosterone data, $n$ = 107 individuals with stressed corticosterone data), and we only measured telomere length in approximately half of nestlings ($n$ = 55 individuals with telomere data). In the model predicting inter-annual return, we included only nestlings that successfully fledged. Models testing the effects of corticosterone treatment on variables measured with radio tag data included only 35 individuals that fledged with radio tags (one of the females that fledged with a radio tag was excluded because of uncertainty surrounding her fledging date). The sample size of nestlings and the number of nests from which they came, for each model, are reported in the appendix, available online.

    To obtain an estimate of how well our best-fit models fit our data, we calculated $R^2$ for our best-fit models using the function "r.squaredGLMM" in the MuMIn package (Bartón 2016) in R. This function uses the methods for calculating $R^2$ for mixed-effect models presented in Nakagawa et al. (2013) and provides a marginal $R^2$ estimate ($R^2m$), which uses only fixed effects, and a conditional $R^2$ estimate ($R^2c$), which takes both fixed and random effects into account. The output data from each best-fit model are summarized in the appendix, available online.

**Results**

Model selection results are summarized in Table 1. The best fit model of nestling body mass suggests that control individuals were heavier than corticosterone-treated individuals, and that males were heavier than females (Figure 1; $R^2m = 0.13$, $R^2c = 0.59$, weight = 0.486; treatment (control) β = 0.69, SE = 0.33, df = 76, $P = 0.041$; sex (male) β = 1.57, SE = 0.34, df = 76, $P < 0.001$). Scaled body mass showed a similar pattern ($R^2m = 0.082$, $R^2c = 0.57$, weight = 0.468; treatment (control) β = 0.70, SE = 0.28, df = 74, $P = 0.015$; sex (male) β = 0.85, SE = 0.29, df = 74, $P = 0.004$). The best model of head-bill was the null model (weight = 0.372). Wing length was best predicted by the model with sex, brood size, and hatch date ($R^2m = 0.16$, $R^2c = 0.64$, weight = 0.384; sex (male) β = 1.20, SE = 0.95, df = 77, $P = 0.21$; brood size β = -1.93, SE = 1.29, df = 18, $P = 0.15$; hatch date β = 0.79, SE = 0.38, df = 18, $P = 0.05$)

Both baseline and stress-induced corticosterone were higher in treatment individuals (Figures 2, 3). The mean corticosterone concentration of control birds was 7.6 ± 14.9 ng mL-1 (baseline, n = 55) and 19.5 ± 13.4 ng mL-1 (stress-induced, n = 53); the mean corticosterone concentration for treated birds (measured 2 days after the last dose was applied) was 21.4 ± 20.8 ng mL-1 (baseline, n = 55) and 38.1 ± 33.6 ng mL-1 (stress-induced, n = 54). Baseline corticosterone was higher in males, individuals from small broods, and earlier-hatching individuals ($R^2m = 0.16$, $R^2c = 0.16$, weight = 0.390; treatment (control) β = -11.43, SE = 3.42, df = 76, $P = 0.001$; sex (male) β = 1.76, SE = 3.46, df = 76, $P = 0.61$; brood size β = -4.07, SE = 2.05, df = 18, $P = 0.062$; hatch date β = -0.98, SE = 0.56, df = 18, $P = 0.097$). The best-fit model for stress-induced corticosterone contained an interaction between treatment and sex: although both sexes showed increased stress-induced corticosterone in response to the treatment, the effect was stronger in females ($R^2m = 0.15$, $R^2c = 0.23$, weight = 0.429; treatment (control) β = -25.27,

SE = 6.87, df = 74, P <0.001; sex (male) β = -11.66, SE = 6.29, df = 74, $P$ = 0.070; interaction (control:male) β = 16.97, SE = 9.29, df = 74, $P$ = 0.072). The estimates for linear and mixed-effect models were identical in the model for baseline corticosterone and similar for stress-induced corticosterone, and we used a likelihood ratio test between our best-fit mixed-effect model for each of these variables and the corresponding linear model. The results did not support the inclusion of the random effect in either case (LRT P > 0.05 for both baseline and stress-induced cort).

The best-fit model of telomere length suggests that the corticosterone treatment reduced telomere length ($R^2m$ = 0.028, $R^2c$ = 0.67, weight = 0.547; treatment (control) β = 0.40, SE = 0.19, df = 42, $P$ = 0.04).

We did not find evidence that our corticosterone treatment affected the age at which juveniles departed from the nest or departed from the natal site for the first time: the best-fit models for each of these response variables were null models (with weights of 0.653 and 0.671, respectively). We did find a significant positive interaction between treatment and sex on the number of days individuals spent at the natal site after fledging (Figure 4; $R^2m$ = 0.27, $R^2c$ = 0.50, weight = 0.363; treatment (control) β = 12.26, SE = 3.88, df = 11, $P$ = 0.009; sex (male) β = 3.18, SE = 3.05, df = 11, $P$ = 0.32; interaction (control:male) β = -11.71, SE = 4.33, df = 11, $P$ = 0.02). However, this effect is driven by a single control female that spent an unusually long time at the natal site (25 days); running the analysis without this individual resulted in a lowest AIC score for the null model (weight = 0.56). Most other individuals of both sexes and treatment groups spent less than 10 days at the site post-fledging (Figure 4.)

The best fit model for survival to fledging was the null model, with a weight of 0.368. The best fit model of inter-annual return rate showed a negative effect of radio tagging on return rate and also included a positive effect of corticosterone treatment on return ($R^2m$ = 0.0006, $R^2c$ = 0.0006, weight = 0.379; treatment (control) β = -1.03, SE = 0.74, $P$ = 0.17; tag status (true) β = -1.97, SE = 1.08, $P$ = 0.069). Because we could not detect birds using the radio tags in 2016, we could not distinguish dispersal from mortality among individuals that failed to return to the natal site the year after they hatched.

**Discussion**

 Our results demonstrate that, in this wild population of songbirds, a small number of short-term elevations in corticosterone during development can affect multiple components of phenotype; however, we found no influence of our treatment on survival to fledging or the age at which nestlings departed from the nest or from the natal site, and no negative influence of our treatment inter-annual return rate. Corticosterone-treated nestlings had lower body mass and were in poorer condition than controls (Figure 1), but did not differ in structural size. Corticosterone-treated nestlings also had higher baseline and stress-induced corticosterone levels than controls (Figures 2, 3); but these elevated corticosterone levels were still within the normal physiological range for nestlings in this population (Stedman et al., 2017). In all of these cases, all models within 2 AICc units of the best-fit model also included treatment as a predictor. Our results follow the same patterns shown by previous studies (Loiseau et al. 2008b, Spencer and Verhulst 2007, and Crino et al. 2014); but in these earlier studies, corticosterone dosing was more intensive or of longer duration than in the present study, and experimental manipulations were conducted in captive populations.

Our results suggest that males and females were affected differently by our treatment. One control female spent many more days at the natal site after fledging than any other individual (Figure 4), resulting in a significant relationship between sex, treatment, and days spent at the site; but greater sample sizes are needed to know whether this reflects a true biological pattern or a statistical anomaly. Studies that assess sex-based differences in response to developmental stress often find significant sex differences, but in birds, neither sex has emerged as consistently more sensitive. Males show greater sensitivity in some studies (e.g.,

Love et al. 2005, Spencer and Verhulst 2007) and females in others (e.g., Verhulst et al. 2006, Marasco et al. 2012, this study); additional study is needed to dissect these patterns.

Our results suggest that corticosterone treatment shortened telomeres. However, the effect appears to be weak: the treatment predictor explained only ~3% of telomere length variation ($R^2m = 0.028$). Previous work has also shown a negative effect of corticosterone levels on telomere length (Haussmann et al. 2012, Herborn et al. 2014), which may influence life history by impacting the rate of senescence (e.g., Haussmann and Mauck 2008, Dantzer and Fletcher 2015). The complex relationships among hormones, telomeres, and life history remain poorly understood, and are important targets for further study (Belmaker 2016, Haussmann and Heidinger 2015).

Although fledging and natal site departure have been correlated with spikes in corticosterone (Sprague and Breuner 2010, Belthoff and Dufty 1998), we did not find evidence that a corticosterone treatment applied during the earlier nestling period affected age at fledging or age at initial departure from the natal site. There may be no strong causal link between stress experienced earlier in development and the timing of these major life transitions. Other factors possibly affecting fledging age in birds include social cues from siblings, and parental behavior (Michaud and Leonard 2000, Deguchi et al. 2004).

We also found no clear link between corticosterone treatment and survival to fledging. In general, the effects of corticosterone on survival are likely to be context-dependent; a consistent relationship between corticosterone and survival has not yet emerged (e.g. Boonstra 2013, Crespi et al 2013). Our experiment took place in a particularly good year for tree swallows: 80% of nestlings involved in our experiment fledged, compared to a mean nestling fledge rate of 58% at our site between 2013 and 2016. It is possible that in poor years - i.e. years with lower spring temperatures, differing food availability, and subsequently higher mortality - corticosterone manipulation would impact survival to fledging. It is also possible that nestling corticosterone levels would more strongly affect survival after fledging (when young swallows are becoming independent and challenged by new environments and the need to catch their own food) than in the relatively-stable environment of the nest; a large proportion of mortality in hatch-year swallows occurs during the few weeks immediately following fledging (Grüebler et al. 2014). However, we also found no evidence for a negative effect of corticosterone on inter-annual return rate (a metric that reflects both annual survival and dispersal decision.) Radio tagging had a negative effect on return rate. Treatment was included as a predictor in the best model predicting return rate, but the effect of corticosterone on survival in this model was slightly positive.

Overall, we found strong evidence that corticosterone treatment affects morphology and baseline and stress-induced corticosterone in our study; but the fixed-effect predictors in our best fit models tended to explain only a small portion of the variation we saw. Differences between marginal and conditional $R^2$ estimates show that much of the variation explained by some of our models is explained by our random effect, which was the nest each nestling came from. Marginal $R^2$, which takes only the fixed effects and not the random effects into account, ranged from essentially 0 (for the model predicting inter-annual return rate) to 0.27 (for the model predicting total days present at the natal site after fledging). Conditional $R^2$, which accounts for random effects as well as fixed effects, was much higher (up to 0.67), particularly in the models predicting morphology, telomere length, and total days present at the natal site after fledging. Interestingly, conditional $R^2$ was not much higher than marginal $R^2$ for both baseline and stress-induced corticosterone and a likelihood ratio test did not support the inclusion of the random effect in either model. In other words, while a great deal of variation in morphological and telomere measurements is explained by the nest a particular nestling comes from, the same cannot be said of corticosterone levels. Previous research on this population (Stedman et al. 2017) has shown significant effects of both nest environment and genetic background on variation in corticosterone, suggesting that our manipulation of the nestlings may have disrupted or overwhelmed the effect of rearing environment on corticosterone physiology.

The harshness of the nestling environment at our experimental site varies from year to year with weather. During periods of cold weather, food availability is greatly decreased and parents may leave their nests for unusually prolonged periods of time; widespread nestling mortality occurs in colder years (Winkler et al. 2013). Our results suggest that if variation in the nestling environment (including cold weather and other stressors) causes short-term stress, this variation can alter phenotype via the HPA axis: these short-term stressors may have a greater impact than previously recognized, even if their effects do not persist over the long term. It may be that more challenging breeding seasons, where nestlings have higher glucocorticoid levels, could produce generations of nestlings with corresponding differences in morphology and stress responsiveness. Future studies could address whether the developmental environment has long-term impacts on phenotype, and what implications this would have for individual fitness.

**Acknowledgements**

We thank the many undergraduate members of the Unit 1 and Unit 2 Tree Swallow research crews for help in the field. Kelly Hallinger and Bronwyn Butcher provided advice for molecular sexing and Conor Taff provided advice for hormone assays. This work was supported by an NSF LTREB grant (DEB-1242573) to DWW and by an NSF grant (IOS-1457251) to MNV. All work with animals reported here was conducted under approved Cornell Animal Protocol No. 2001-0051, Federal Master Banding Permit No. 20576, and NY State Banding License No. 106 to DWW.

**Literature Cited**

Bakaysa, S. L., L. A. Mucci, P. E. Slagboom, D. I. Boomsma, G. E. McClearn, B. Johansson, and N. L. Pederson. 2007. Telomere length predicts survival independent of genetic influences. Aging Cell 6:769-774.

Bartón, K. 2016. MuMIn: Multi-Model Inference. R package.

Belmaker, A. 2016. The role of telomere length in the life history and behavior of Tree Swallows (*Tachycineta bicolor*). PhD diss, Cornell University, Ithaca.

Belthoff, J. R. and A. M. Dufty. 1998. Corticosterone, body condition and locomotor activity: a model for dispersal in screech-owls. Anim Behav 55:405-415.

Boonekamp, J. J., G. A. Mulder, H. M. Salomons, C. Dijkstra and S. Verhulst. 2014. Nestling telomere shortening, but not telomere length, reflects developmental stress and predicts survival in wild birds. Proc. Biol. Sci. 281: 20133287.

Boonekamp, J. J., C. Bauch, E. Mulder, and S. Verhulst. 2017. Does oxidative stress shorten telomeres? Biol. Lett. 13: 20170164.

Boonstra, R. 2013. Reality as the leading cause of stress: rethinking the impact of chronic stress in nature. Funct Ecol 27: 11-23.

Breuner, C. W., A. L. Greenberg, and J. C. Wingfield. 1998. Noninvasive corticosterone treatment rapidly increases activity in Gambel's White-crowned Sparrows (*Zonotrichia leucophrys gambelii*). Gen. Comp. Endocrinol. 111:386-394.

Challet, E., Y. Le Maho, J.-P. Robin, A. Malan, and Y. Cherel. 1995. Involvement of corticosterone in the fasting-induced rise in protein utilization and locomotor activity. Pharmacol. Biochem. Behav. 50: 405-412.

Crespi, E. J., T.D. Williams, T.S. Jessop, and B. Delehanty. 2013. Life history and the ecology of stress: how do glucocorticoid hormones influence life-history variation in animals? Funct Ecol, 27: 93-106.

Crino, O. L., S. C. Driscoll, R. Ton and C. W. Breuner. 2014. Corticosterone exposure during development improves performance on a novel foraging task in zebra finches. Anim Behav 91:27-32.

Dantzer, B. and Q. E. Fletcher. 2015. Telomeres shorten more slowly in slow-aging wild animals than in fast-aging ones. Exp. Gerontol. 71:38-47.

Deguchi, T., A. Takahashi, and Y. Watanuki. 2004. Proximate factors determining age and mass at fledging in rhinoceros auklets (Cerorhinca monocerata): intra-and interyear variations. The Auk, 121: 452-462.

Griffiths, R., M. C. Double, K. Orr and J. G. Dawson. 1998. A DNA test to sex most birds. Mol. Ecol. 7:1071-1075.

Grüebler, M. U., F. Korner-Nievergelt and B. Naef-Daenzer. 2014. Equal nonbreeding period survival in adults and juveniles of a long-distant migrant bird. Ecol Evol 4:756-765.

Haussmann, M. F. and B. J. Heidinger. 2015. Telomere dynamics may link stress exposure and ageing across generations. Biol. Lett. 11:20150396.

Haussmann, M. F. and N. M. Marchetto. 2010. Telomeres: Linking stress and survival, ecology and evolution. Curr Zool 56:714-727.

Haussmann, M. F. and R. A. Mauck. 2008. Telomeres and longevity: testing an evolutionary hypothesis. Mol. Biol. and Evol. 25:220-228.

Haussmann, M. F., D. W. Winkler, K. M. O'Reilly, C. E. Huntington, I. C. T. Nisbet, and C. M. Vleck. 2003. Telomeres shorten more slowly in long-lived birds and mammals than in short-lived ones. Proc. Biol. Sci. 270:1387-1392.

Haussmann, M. F., D. W. Winkler and C. M. Vleck. 2005. Longer telomeres associated with higher survival in birds. Biol. Lett. 1:212-214.

Haussmann, M. F., A. S. Longenecker, N. M. Marchetto, S. A. Juliano and R. M. Bowden. 2012. Embryonic exposure to corticosterone modifies the juvenile stress response, oxidative stress and telomere length. Proc. Biol. Sci. 279:1447-1456.

Hayward, L. S. and J. C. Wingfield. 2004. Maternal corticosterone is transferred to avian yolk and may alter offspring growth and adult phenotype. Gen. Comp. Endocrinol. 135:365-371.

Herborn, K. A., B. J. Heidinger, W. Boner, J. C. Noguera, A. Adam, F. Daunt and P. Monaghan 2014. Stress exposure in early post-natal life reduces telomere length: an experimental demonstration in a long-lived seabird. Proc. Biol. Sci. 281:20133151.
26

Jones, J. 2003. Tree Swallows (*Tachycineta bicolor*): A new model organism? The Auk 120:591-599.

Kimura, M., J. V. Hjelmborg, J. P. Gardner, L. Bathum, M. Brimacombe, X. Lu, L. Christiansen, J. W. Vaupel, A. Aviv and K. Christensen. 2008. Telomere length and mortality: a study of leukocytes in elderly Danish twins. Am. J. Epidemiol. 167: 799-806.

Landys, M. M., M. Ramenofsky, and J. C. Wingfield. 2006. Actions of glucocorticoids at a seasonal baseline as compared to stress-related levels in the regulation of periodic life processes. Gen. Comp.Endocrinol. 148:132-149.

Loiseau, C., G. Sorci, S. Dano and O. Chastel. 2008a. Effects of experimental increase of corticosterone levels on begging behavior, immunity and parental provisioning rate in house sparrows. Gen. Comp. Endocrinol. 155:101-108.

Loiseau, C., S. Fellous, C. Haussy, O. Chastel and G. Sorci. 2008b. Condition-dependent effects of corticosterone on a carotenoid-based begging signal in house sparrows. Horm Behav 53:266-273.

Love, O. P., E. H. Chin, K. E. Wynne-Edwards and T. D. Williams. 2005. Stress hormones: A link between maternal condition and sex-biased reproductive investment. Am. Nat. 166:751-766.

Marasco, V., J. Robinson, P. Herzyk and K. A. Spencer. 2012. Pre- and post-natal stress in context: effects on the stress physiology in a precocial bird. J. Exp. Biol. 215:3955-3964.

McCarty, J. P. 2001. Variation in growth of nestling Tree Swallows across multiple temporal and spatial scales. The Auk 118:176-190.

McEwen, B. S. and J. C. Wingfield. 2003. The concept of allostasis in biology and biomedicine. Horm Behav 43:2-15.

Meylan, S., M. De Fraipont, P. Aragon, E. Vercken, and J. Clobert. 2009. Are dispersal-dependent behavioral traits produced by phenotypic plasticity? J Exp Zool A Ecol Integr Physiol 311: 377-88.

Michaud, T., and M. Leonard. 2000. The role of development, parental behavior, and nestmate competition in fledging of nestling Tree Swallows. The Auk 117:996-1002.

Morton, M. L. 1992. Effects of sex and birth date on premigration biology, migration schedules, return rates, and natal dispersal in the Mountain White-crowned Sparrow. Condor 94:117-133.

Nakagawa, S., H. Schielzeth and R. B. O'Hara. 2013. A general and simple method for obtainingR2from generalized linear mixed-effects models. Methods Ecol Evol 4:133-142.

Peig, J. and A. J. Green. 2009. New perspectives for estimating body condition from mass/length data: the scaled mass index as an alternative method. OIKOS 118:1883-1891.

R Core Team. 2017. R: a language and environment for statistical computing. R Foundation for Statistical Computing, Vienna. http://www.R-project.org/.

Rappole, J. H., and A. R. Tipton. 1991. New harness design for attachment of radio transmitters to small passerines. J Field Ornithol 62:335-337.

Reed, J. M., T. Boulinier, E. Danchin, and L. W. Oring. 1999. Informed dispersal: prospecting by birds for breeding sites. Pp. 189-259. In: V. Nolan, Jr., ed. Current Ornithology, Vol 15. Kluwer Academic/Plenum Publishers, New York, NY.

Rivers, J. W., A. L. Liebl, J. C. Owen, L. B. Martin, M. G. Betts, and J. Grindstaff. 2012. Baseline corticosterone is positively related to juvenile survival in a migrant passerine bird. Funct Ecol 26:1127-1134.

Safran, R. J., C. R. Neuman, K. J. McGraw and I. J. Lovette. 2005. Dynamic paternity allocation as a function of male plumage color in barn swallows. Science 309:2210-2212.

Sapolsky, R. M., L. M. Romero and A. U. Munck. 2000. How do glucocorticoids influence stress responses? Integrating permissive, suppressive, stimulatory, and preparative actions. Endocr. Rev. 21:55-89.

Schmidt, K. L., E. A. Macdougall-Shackleton, K. K. Soma and S. A. Macdougall-Shackleton. 2014. Developmental programming of the HPA and HPG axes by early-life stress in male and female song sparrows. Gen. Comp.Endocrinol. 196:72-80.

Schoech, S. J., M. A. Rensel and R. S. Heiss. 2011. Short- and long-term effects of developmental corticosterone exposure on avian physiology, behavioral phenotype, cognition, and fitness: A review. Curr Zool 57:514-530.

Silverin, B. 1997. The stress response and autumn dispersal behavior in willow tits. Anim Behav 53:451-459.

Spencer, K. A. and S. Verhulst. 2007. Delayed behavioral effects of postnatal exposure to corticosterone in the zebra finch (*Taeniopygia guttata*). Horm Behav 51:273-280.

Spencer, K. A. and S. Verhulst. 2008. Post-natal exposure to corticosterone affects standard metabolic rate in the zebra finch (*Taeniopygia guttata*). Gen. Comp. Endocrinol. 159:250-256.

Sprague, R. S. and C. W. Breuner. 2010. Timing of fledging is influenced by glucocorticoid physiology in Laysan Albatross chicks. Horm Behav 58:297-305.

Stedman, J. M., K. K. Hallinger, D. W. Winkler, and M. N. Vitousek. 2017. Heritable variation in circulating glucocorticoids and endocrine flexibility in a free-living songbird. J. Evol. Biol. 30:1724-1735.

Vercken, E., M. de Fraipont, A. M. Duffy Jr., and J. Clobert. 2007. Mother's timing and duration of corticosterone exposure modulate offspring size and natal dispersal in the common lizard (*Lacerta vivipara*). Horm Behav 51:379-386.

Verhulst, S., M. J. Holveck and K. Riebel. 2006. Long-term effects of manipulated natal brood size on metabolic rate in zebra finches. Biol. Lett. 2:478-480.

Verspoor, J. J., O. P. Love, E. Rowland, E. H. Chin and T. D. Williams. 2007. Sex-specific development of avian flight performance under experimentally altered rearing conditions. Behav. Ecol. 18:967-973.
30

Vitousek, M. N., C. C. Taff, D. R. Ardia, J. M. Stedman, C. Zimmer, T. C. Salzman, and D. W. Winkler. 2018. The lingering impact of stress: brief acute glucocorticoid exposure has sustained, dose-dependent effects on reproduction. Proc. Biol. Sci. 285: 20180722.

Wada, H. and C. W. Breuner. 2008. Transient elevation of corticosterone alters begging behavior and growth of white-crowned sparrow nestlings. J. Exp. Biol. 211:1696-1703.

Whittingham, L. A. and P. O. Dunn. 2000. Offspring sex ratios in tree swallows: females in better condition produce more sons. Mol. Ecol. 9:1123-1129.

Winkler, D. W. and P. E. Allen. 1996. The seasonal decline in Tree Swallow clutch size: physiological constraint or strategic adjustment? Ecology 77:922-932.

Winkler, D. W., P. H. Wrege, P. E. Allen, T. L. Kast, P. Senesac, M. F. Wasson and P. J. Sullivan. 2005. The natal dispersal of tree swallows in a continuous mainland environment. J Anim Ecol 74:1080-1090.

Winkler, D. W., M. K. Luo and E. Rakhimberdiev. 2013. Temperature effects on food supply and chick mortality in tree swallows (*Tachycineta bicolor*). Oecologia 173:129-138.

**Tables**

Table 1. A summary of results for models within 2 ∂AICc of the best-fit model.

| Response variable | Best-fit models | ∂AICc from best-fit model | Weight |
|---|---|---|---|
| Mass | 1. treatment + sex | 0 | 0.49 |
| | 2. treatment + sex + brood size + hatch date | 1.05 | 0.22 |
| | 3. treatment + sex + treatment:sex | 1.38 | 0.21 |
| Head-bill | 1. null | 0 | 0.37 |
| | 2. treatment + sex | 0.53 | 0.24 |
| | 3. treatment | 1.19 | 0.19 |
| Wing length | 1. sex + brood size + hatch date | 0 | 0.38 |
| | 2. null | 0.80 | 0.26 |
| | 3. treatment + sex + brood size + hatch date | 1.90 | 0.15 |
| Scaled body mass | 1. treatment + sex | 0 | 0.47 |
| | 2. treatment + sex + brood size + hatch date | 0.90 | 0.30 |
| | 3. treatment + sex + treatment:sex | 1.99 | 0.17 |
| Baseline cort | 1. treatment + sex + brood.size + hatch date | 0 | 0.39 |
| | 2. treatment | 0.82 | 0.39 |
| Stress-induced cort | 1. treatment + sex + treatment:sex | 0 | 0.43 |
| | 2. treatment | 0.05 | 0.34 |
| | 3. treatment + sex | 1.36 | 0.20 |
| Telomere length | 1. treatment | 0 | 0.55 |
| | 2. treatment + sex | 1.96 | 0.21 |
| Age at fledging | 1. null | 0 | 0.65 |
| | 2. treatment | 1.88 | 0.19 |
| Age at first departure from natal site | 1. null | 0 | 0.67 |
| | 2. treatment | 1.95 | 0.19 |
| Total days at the natal site after fledging | 1. treatment + sex + treatment:sex | 0 | 0.36 |
| Survival to fledge | 1. null | 0 | 0.37 |
| | 2. treatment + tag status | 0.015 | 0.37 |
| | 3. tag status | 1.74 | 0.15 |
| Inter-annual | 1. treatment+tag status | 0 | 0.38 |

| return | 2. tag status | 0.055 | 0.35 |

**Figure Legends**

**Figure 1.** Boxplots showing the difference in nestling body mass on day 12 by treatment (A) and sex (B). Numbers in boxes indicate sample sizes.

**Figure 2.** Boxplots showing the difference in baseline corticosterone (as measured on day 12) by treatment. Numbers indicate sample sizes.

**Figure 3.** Boxplots showing the relationship between treatment, sex, and stress-induced corticosterone (as measured on day 12). Numbers in boxes indicate sample sizes. The plot demonstrates the interaction between sex and treatment by dividing the sample of nestlings into 4 groups based on their treatment and sex.

**Figure 4.** Boxplots showing the relationship between treatment, sex, and the number of days spent at the natal site after fledging. Numbers in boxes indicate sample sizes. The plot demonstrates the interaction between sex and treatment by dividing the sample of nestlings into 4 groups based on their treatment and sex.

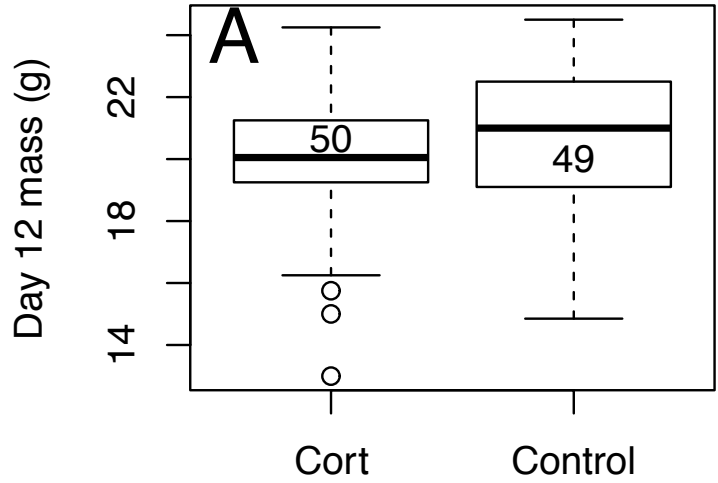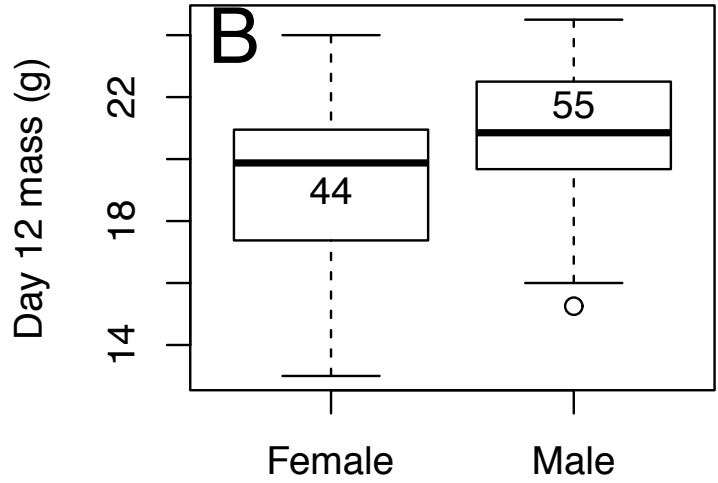

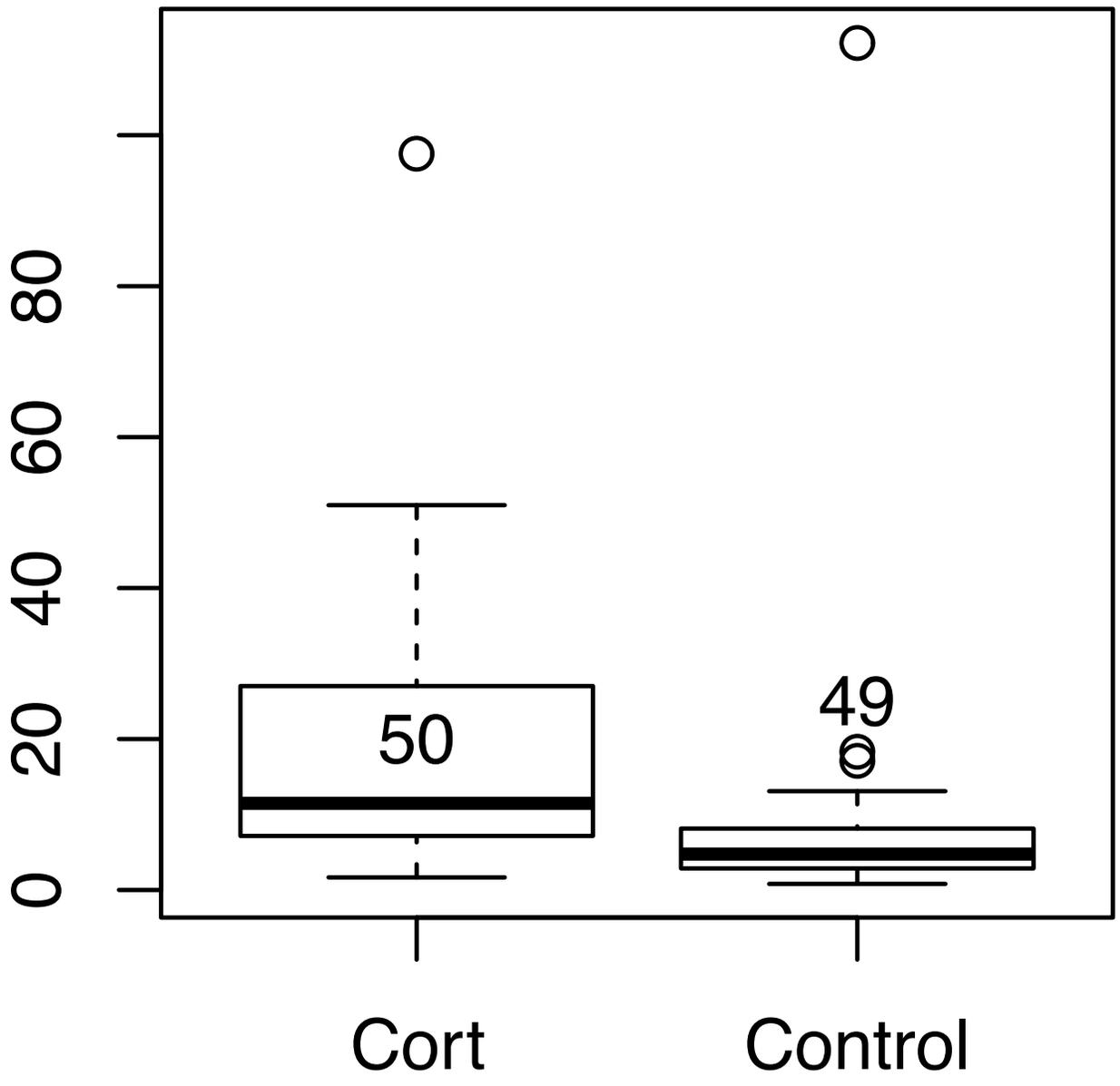

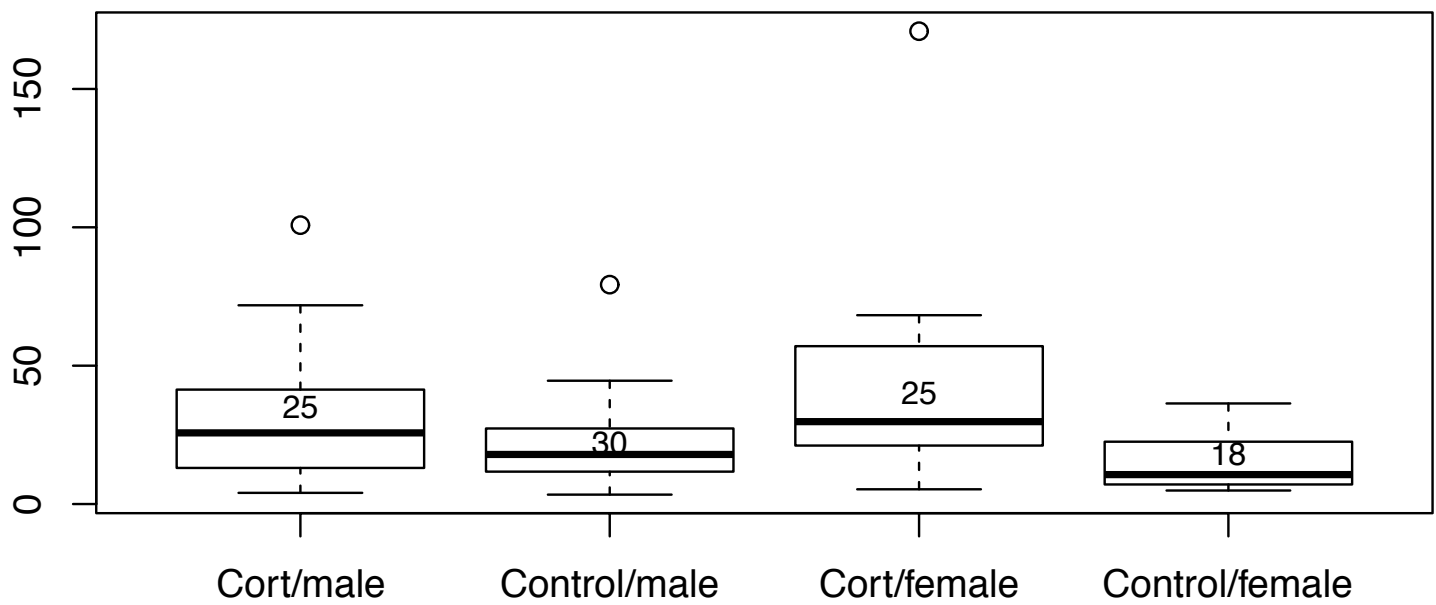

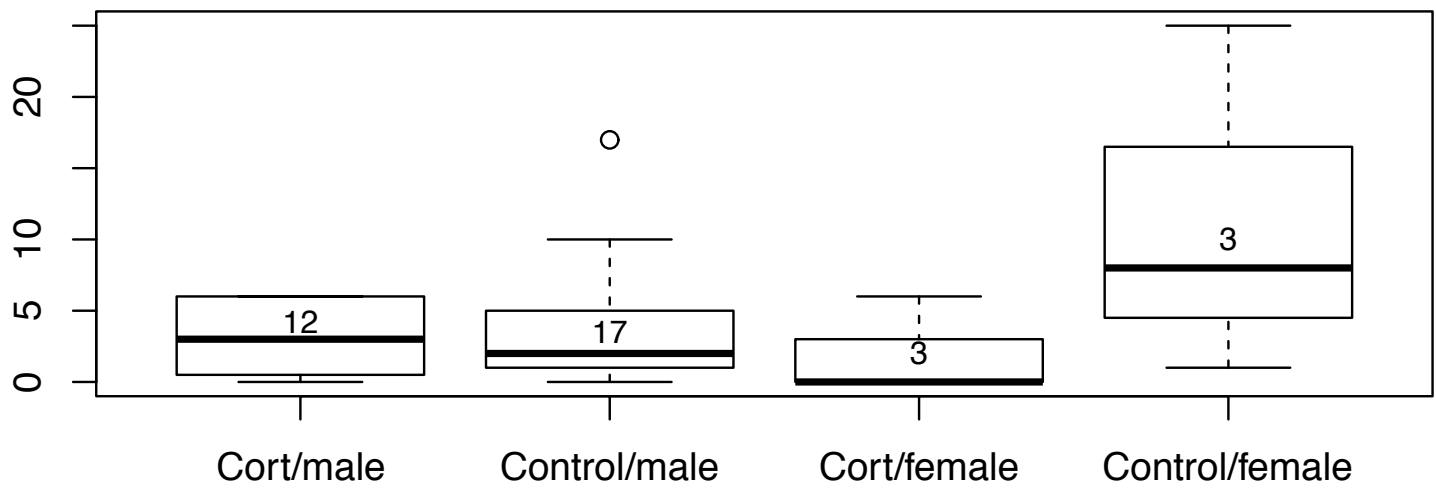

APPENDIX

Detailed results of model selection

Table S1. Model selection results for models predicting day 12 mass. The AICc value of the best-fit model was 420.12. Data for these models include 99 individuals from 21 nests.

| Predictors | Log Likelihood | $\partial$AICc | Weight |
|---|---|---|---|
| treatment + sex | -205.11 | 0 | 0.486 |
| treatment + sex + brood size + hatch date | -203.63 | 1.047 | 0.215 |
| treatment + sex + treatment:sex | -204.80 | 1.380 | 0.213 |
| sex + brood size + hatch date | -205.72 | 3.223 | 0.085 |
| treatment | -214.75 | 17.285 | 0.0001 |
| null | -217.38 | 20.558 | 0.00002 |

Table S2. Best-fit model details for model predicting day 12 mass. For this model, $R^2m = 0.13$ and $R^2c = 0.59$.

| | Value | Standard Error | DF | t-value | P-value |
|---|---|---|---|---|---|
| **intercept** | 19.01 | 0.471 | 76 | 40.304 | 0 |
| **treatment (control)** | 0.689 | 0.332 | 76 | 2.075 | 0.0414 |
| **sex (male)** | 1.575 | 0.341 | 76 | 4.616 | 0 |

Table S3. Model selection results for models predicting day 12 headbill. The AICc value of the best-fit model was 295.23. Data for these models include 99 individuals from 21 nests.

| Predictors | Log Likelihood | ∂AICc | Weight |
|---|---|---|---|
| null | -144.61 | 0 | 0.372 |
| treatment + sex | -142.88 | 0.529 | 0.235 |
| treatment | -144.21 | 1.186 | 0.189 |
| treatment + sex + treatment:sex | -142.76 | 2.291 | 0.085 |
| sex + brood size + hatch date | -142.88 | 2.533 | 0.075 |
| treatment + sex + brood size + hatch date | -142.27 | 3.318 | 0.043 |

Table S4. Best-fit model details for model predicting day 12 head-bill.

| | Value | Standard Error | DF | t-value | P-value |
|---|---|---|---|---|---|
| **intercept** | 25.366 | 0.199 | 78 | 127.528 | 0 |

Table S5. Model selection results for models predicting day 12 flat wing. The AICc value of the best-fit model was 628.12. Data for these models include 99 individuals from 21 nests.

| Predictors | Log Likelihood | ∂AICc | Weight |
|---|---|---|---|
| sex + brood size + hatch date | -308.06 | 0 | 0.384 |
| null | -311.46 | 0.80 | 0.257 |

| Predictors | Log Likelihood | ∂AICc | Weight |
|---|---|---|---|
| treatment + sex + brood size + hatch date | -308.01 | 1.895 | 0.149 |
| treatment | -311.34 | 2.567 | 0.106 |
| treatment + sex | -310.69 | 3.272 | 0.075 |
| treatment + sex + treatment:sex | -310.66 | 5.199 | 0.029 |

Table S6. Best-fit model details for model predicting day 12 flat wing. For this model, $R^2m$ = 0.16 and $R^2c$ = 0.64.

| | Value | Standard Error | DF | t-value | P-value |
|---|---|---|---|---|---|
| **intercept** | -57.88 | 57.211 | 77 | -1.012 | 0.315 |
| **sex (male)** | 1.203 | 0.949 | 77 | 1.268 | 0.209 |
| **brood size** | -1.932 | 1.290 | 18 | -1.498 | 0.152 |
| **hatch date** | 0.789 | 0.384 | 18 | 2.053 | 0.0549 |

Table S7. Model selection results for models predicting day 12 condition. The AICc value of the best-fit model was 378.46. Data for these models include 97 individuals from 21 nests.

| Predictors | Log Likelihood | ∂AICc | Weight |
|---|---|---|---|
| treatment + sex | -184.23 | 0 | 0.468 |
| treatment + sex + brood size + hatch date | -182.68 | 0.879 | 0.299 |
| treatment + sex + treatment:sex | -184.23 | 1.989 | 0.173 |
| sex + brood size + hatch date | -185.73 | 4.991 | 0.039 |

| Predictors | Log Likelihood | ∂AICc | Weight |
|---|---|---|---|
| treatment | -188.41 | 6.365 | 0.019 |
| null | -192.02 | 11.568 | 0.001 |

Table S8. Best-fit model details for model predicting day 12 scaled body mass (condition). For this model, $R^2m = 0.082$, and $R^2c = 0.57$.

|  | Value | Standard Error | DF | t-value | P-value |
|---|---|---|---|---|---|
| **intercept** | 19.867 | 0.401 | 74 | 49.501 | 0 |
| **treatment (control)** | 0.700 | 0.282 | 74 | 2.482 | 0.0153 |
| **sex (male)** | 0.850 | 0.290 | 74 | 2.939 | 0.0044 |

Table S9. Model selection results for models predicting day 12 baseline cort. The AICc value of the best-fit model was 849.71. Data for these models include 99 individuals from 21 nests.

| Predictors | Log Likelihood | ∂AICc | Weight |
|---|---|---|---|
| treatment + sex + brood size + hatch date | -417.86 | 0 | 0.390 |
| treatment | -421.27 | 0.824 | 0.386 |
| treatment + sex | -421.17 | 2.630 | 0.140 |
| treatment + sex + treatment:sex | -420.72 | 3.735 | 0.071 |
| null | -426.14 | 8.571 | 0.009 |

| Predictors | Log Likelihood | ∂AICc | Weight |
|---|---|---|---|
| sex + brood size + hatch date | -423.40 | 9.095 | 0.005 |

Table S10. Best-fit model details for model predicting day 12 baseline cort. For this model, $R^2m$ = 0.16 and $R^2c$ = 0.16.

| | Value | Standard Error | DF | t-value | P-value |
|---|---|---|---|---|---|
| **intercept** | 186.249 | 84.783 | 76 | 2.197 | 0.0311 |
| **treatment (control)** | -11.429 | 3.423 | 76 | -3.339 | 0.0013 |
| **sex (male)** | 1.759 | 3.459 | 76 | 0.508 | 0.6126 |
| **brood size** | -4.069 | 2.045 | 18 | -1.990 | 0.0620 |
| **hatch date** | -0.980 | 0.559 | 18 | -1.753 | 0.0967 |

Table S11. Model selection results for models predicting day 12 stressed cort. The AICc value of the best-fit model was 896.34. Data for these models include 98 individuals from 21 nests.

| Predictors | Log Likelihood | ∂AICc | Weight |
|---|---|---|---|
| treatment + sex + treatment:sex | -442.17 | 0 | 0.429 |
| treatment | -444.19 | 0.050 | 0.343 |
| treatment + sex | -443.85 | 1.360 | 0.199 |
| treatment + sex + brood size + hatch date | -443.58 | 4.831 | 0.026 |
| null | -450.52 | 10.696 | 0.002 |
| sex + brood size + hatch date | -449.48 | 14.627 | 0.0002 |

Table S12. Best-fit model details for model predicting day 12 stressed cort. For this model, $R^2m$ = 0.15 and $R^2c$ = 0.23.

|  | Value | Standard Error | DF | t-value | P-value |
|---|---|---|---|---|---|
| **intercept** | 41.643 | 4.656 | 74 | 8.943 | 0 |
| **treatment (control)** | -25.273 | 6.867 | 74 | -3.680 | 0.0004 |
| **sex (male)** | -11.662 | 6.294 | 74 | -1.853 | 0.0679 |
| **treatment (control):sex (male)** | 16.966 | 9.287 | 74 | 1.827 | 0.0718 |

Table S13. Model selection results for models predicting day 12 telomere length. The AICc value of the best-fit model was 150.72. Data for these models include 55 individuals from 12 nests.

| **Predictors** | Log Likelihood | ∂AICc | Weight |
|---|---|---|---|
| treatment | -71.36 | 0 | 0.547 |
| treatment + sex | -71.34 | 1.969 | 0.212 |
| null | -73.47 | 2.225 | 0.165 |
| treatment + sex + treatment:sex | -71.16 | 3.609 | 0.056 |
| treatment + sex + brood size + hatch date | -71.24 | 5.753 | 0.014 |
| sex + brood size + hatch date | -73.36 | 7.998 | 0.006 |

Table S14. Best-fit model details for model predicting day 12 telomere length. For this model, $R^2m = 0.028$ and $R^2c = 0.67$.

|  | Value | Standard Error | DF | t-value | P-value |
|---|---|---|---|---|---|
| **intercept** | 9.447 | 0.315 | 42 | 30.035 | 0 |
| **treatment (control)** | 0.401 | 0.193 | 42 | 2.084 | 0.0433 |

Table S15. Model selection results for models predicting fledge age. The AICc value of the best-fit model was 138.55. Data for these models include 35 individuals from 21 nests.

| Predictors | Log Likelihood | ∂AICc | Weight |
|---|---|---|---|
| null | -66.27 | 0 | 0.653 |
| treatment | -66.22 | 1.884 | 0.192 |
| sex + brood size + hatch date | -64.31 | 2.064 | 0.076 |
| treatment + sex + brood size + hatch date | -64.21 | 3.868 | 0.049 |
| treatment + sex | -66.21 | 3.874 | 0.017 |
| treatment + sex + treatment:sex | -66.19 | 5.838 | 0.012 |

Table S16. Best-fit model details for model predicting fledge age.

|  | Value | Standard Error | DF | t-value | P-value |
|---|---|---|---|---|---|
| **intercept** | 20.27 | 0.304 | 21 | 66.711 | 0 |

Table S17. Model selection results for models predicting age at first departure from the natal site. The AICc value of the best-fit model was 120.60. Data for these models include 35 individuals from 21 nests.

| Predictors | Log Likelihood | ∂AICc | Weight |
|---|---|---|---|
| null | -57.14 | 0 | 0.671 |
| treatment | -57.11 | 1.947 | 0.192 |
| treatment + sex + treatment:sex | -55.84 | 3.395 | 0.055 |
| sex + brood size + hatch date | -55.98 | 3.698 | 0.040 |
| treatment + sex | -57.00 | 3.725 | 0.035 |
| treatment + sex + brood size + hatch date | -55.95 | 5.630 | 0.007 |

Table S18. Best-fit model details for model predicting age at first departure from the natal site.

|  | Value | Standard Error | DF | t-value | P-value |
|---|---|---|---|---|---|
| **intercept** | 0.950 | 0.248 | 21 | 3.829 | 0.001 |

Table S19. Model selection results for models predicting total days spent at the natal site after fledging. The AICc value of the best-fit model was 216.87. Data for these models include 35 individuals from 21 nests.

| Predictors | Log Likelihood | ∂AICc | Weight |
|---|---|---|---|

| Predictors | Log Likelihood | ∂AICc | Weight |
|---|---|---|---|
| treatment + sex + treatment:sex | -102.43 | 0 | 0.363 |
| null | -106.74 | 2.604 | 0.300 |
| treatment + sex | -104.88 | 2.896 | 0.167 |
| treatment | -106.04 | 3.219 | 0.136 |
| sex + brood size + hatch date | -105.21 | 5.548 | 0.023 |
| treatment + sex + brood size + hatch date | -104.25 | 5.632 | 0.012 |

Table S20. Best-fit model details for model predicting total days spent at the natal site after fledging. For this model, $R^2m = 0.27$ and $R^2c = 0.50$.

|  | Value | Standard Error | DF | t-value | P-value |
|---|---|---|---|---|---|
| **intercept** | -0.109 | 2.742 | 20 | -0.0399 | 0.969 |
| **treatment (control)** | 12.26 | 3.876 | 11 | 3.164 | 0.0090 |
| **sex (male)** | 3.18 | 3.045 | 11 | 1.044 | 0.319 |
| **treatment (control):sex (male)** | -11.71 | 4.327 | 11 | -2.707 | 0.0204 |

Table S21. Model selection results for models predicting survival from day 5 to fledging. The AICc value of the best-fit model was 95.1. Data for these models include 108 individuals from 22 nests.

| Predictors | Log Likelihood | ∂AICc | Weight |
|---|---|---|---|
| null | -45.57 | 0 | 0.368 |
| treatment + tag status | -43.58 | 0.015 | 0.365 |
| tag status | -45.44 | 1.739 | 0.154 |
| treatment + brood size + hatch date + tag status | -43.10 | 3.064 | 0.079 |
| brood size + hatch date + tag status | -44.96 | 4.778 | 0.034 |

Table S22. Best-fit model details for model predicting survival from day 5 to fledging.

| | Value | Standard Error | z-value | P-value |
|---|---|---|---|---|
| **intercept** | 2.033 | 0.456 | 4.457 | 0 |

Table S23. Model selection results for models predicting detection in 2016. The AICc value of the best-fit model was 63.1. Data for these models include 88 individuals from 21 nests.

| Predictors | Log Likelihood | ∂AICc | Weight |
|---|---|---|---|
| treatment + tag status | -27.50 | 0 | 0.379 |
| tag status | -28.53 | 0.055 | 0.354 |
| treatment + sex | -27.50 | 2.000 | 0.115 |
| null | -31.16 | 3.313 | 0.080 |
| treatment + sex + treatment:sex + tag status | -27.40 | 3.793 | 0.040 |

| Predictors | Log Likelihood | ∂AICc | Weight |
|---|---|---|---|
| treatment + sex + brood size + hatch date + tag status | -27.21 | 5.415 | 0.017 |
| sex + brood size + hatch date + tag status | -28.24 | 5.488 | 0.015 |

Table S24. Best-fit model details for model predicting detection in 2016. For this model, $R^2m$ = 0.0006 and $R^2c$ = 0.0006.

|  | Value | Standard Error | z-value | P-value |
|---|---|---|---|---|
| **intercept** | -1.137 | 0.448 | -2.536 | 0.0112 |
| **treatment (control)** | -1.025 | 0.743 | -1.379 | 0.168 |
| **tag status (true)** | -1.971 | 1.084 | -1.819 | 0.0689 |